\documentclass[12pt]{article}
\usepackage{amstex}
\begin{document} 
\title{Nambu tensors and commuting vector fields}
\author{J. Hietarinta\\ Department of Physics, University of Turku\\ 
FIN-20014 Turku, Finland\\
email: hietarin@@utu.fi\\ \ }
\date{\today}
\maketitle

\begin{abstract}
Takhtajan has recently studied the consistency conditions for Nambu
brackets, and suggested that they have to be skew-symmetric, and
satisfy Leibnitz rule and the Fundamental Identity (FI, it is a
generalization of the Jacobi identity). If the $n$-th order Nambu
brackets in dimension $N$ is written as $\{f_1,\dots,f_n\}=
\eta_{i_1\dots i_n}\partial_{i_1}f_1\cdots \partial_{i_n}f_n$ (where
the $i_\alpha$ summations range over $1\dots N$), the FI implies two
conditions on the Nambu tensor $\eta$, one algebraic and one
differential.  The algebraic part of FI implies decomposability of
$\eta$ and in this letter we show that the Nambu bracket can then be
written as $\{f_1,\dots,f_n\}=\rho\, \epsilon_{\alpha_1\dots
\alpha_n}\bar D^{{\alpha_1}} f_1\cdots \bar D^{{\alpha_n}}f_n$, where
$\epsilon_{\alpha_1\dots\alpha_n}$ is the usual totally antisymmetric
$n$-dimensional tensor, the $\alpha_i$ summations range over $1 \dots
n$, and $\bar D^{\alpha}:= \partial_\alpha +\sum_{k=n+1}^N
v_k^{\alpha}\partial_k$ are $n$ vector fields. Our main result is that
the differential part of the FI is satisfied iff the vector fields
$\bar D$ commute. Examples are provided by integrable Hamiltonian
systems. It turns out that then the Nambu bracket itself guarantees
that the motions stays on the manifold defined by the constants of
motion of the integrable system, while the $n-1$ Nambu Hamiltonians
determine the (possibly non-integrable) motion on this manifold.
\end{abstract}

\newpage

\section{Introduction}
The standard formulation of Hamiltonian motion using Poisson
brackets is by
\[
\frac{dF}{dt}=\{H,F\},\quad
\{f,g\}:=\frac{\partial(f,g)}{\partial(p,q)}.
\]
In 1973 Nambu proposed an intriguing generalization of this
\cite{N}; the idea was to extend the above classical Poisson bracket
formulation in ${\Bbb R}^2$ to ${\Bbb R}^3$ by generalizing the
Jacobian:
\[
\frac{dF}{dt}=\{H_1,H_2,F\},\quad
\{f,g,h\}:=\frac{\partial(f,g,h)}{\partial(x,y,z)}.
\]
Note the appearance on two Hamiltonians, $H_1$ and $H_2$.
Subsequently Nambu's idea has been extended further, to higher
dimensions (number of free variables), to higher order (number of
functions in the bracket), and to other antisymmetric combinations
than the Jacobian.

Recent interest in this topic is due to Takhtajan \cite{LT}, who
studied in particular the consistency requirements one should place to
such a generalizations; a natural set of properties for the bracket is
\begin{enumerate}
\item Skew symmetry: 
\[
\{f_1,\dots,f_n\}=(-1)^{\epsilon(\sigma)}\{f_{\sigma(1)},
\dots,f_{\sigma(n)}\}
\]
where $\sigma$ is a permutation of $1,\dots,n$ and $\epsilon(\sigma)$
is its parity.
\item Leibnitz rule:
\[
\{ab,f_2,\dots,f_n\}=b\{a,f_2,\dots,f_n\}+a\{b,f_2,\dots,f_n\}.
\]
\item A generalization of Jacobi identity, the Fundamental Identity
(FI) (see also \cite{SV})
\begin{multline}
\{\{h_1,\dots,h_{n-1},f_1\},f_2,\dots,f_n\}+
\{f_1,\{h_1,\dots,h_{n-1},f_2\},f_3,\dots,f_n\}+\cdots\\+
\{f_1,\dots,f_{n-1},\{h_1,\dots,h_{n-1},f_n\}\}=
\{h_1,\dots,h_{n-1},\{f_1,\dots,f_n\}\}.
\end{multline}
\end{enumerate}

If we write the Nambu bracket in terms of the antisymmetric Nambu
tensor $\eta$ \cite{LT}
\begin{equation}
\{f_1,\dots,f_n\}:=\eta_{i_1\dots i_n}\partial_{i_1}f_1\cdots
\partial_{i_n}f_n
\label{E:brad}
\end{equation}
then from the FI it follows \cite{LT} that the Nambu tensor $\eta$
must satisfy two conditions, one algebraic
\begin{equation}
{\cal N}_{i_1i_2\dots i_nj_1j_2\dots j_n}+
{\cal N}_{j_1i_2i_3\dots i_n i_1j_2j_3\dots j_n}=0.
\label{E:ceqn}
\end{equation}
where
\begin{align}
{\cal N}_{i_1i_2\dots i_nj_1j_2\dots j_n}:=
\eta_{i_1i_2\dots i_n}&\eta_{j_1j_2\dots j_n}+
\eta_{j_ni_1i_3\dots i_n}\eta_{j_1j_2\dots j_{n-1} i_2}+
\eta_{j_ni_2i_1i_4\dots i_n}\eta_{j_1j_2\dots j_{n-1} i_3}
\nonumber\\&\cdots+\eta_{j_ni_2i_3\dots i_{n-1}i_1}
\eta_{j_1j_2\dots j_{n-1} i_n}-
\eta_{j_ni_2i_3\dots i_n}\eta_{j_1j_2\dots j_{n-1} i_1}.
\label{E:ceqd}
\end{align}
and one differential \cite{LT}
\begin{align} {\cal D}_{i_2\dots i_nj_1\dots j_n}:=
&\eta_{ki_2\dots i_n}\,\partial_k\,\eta_{j_1j_2\dots j_n}+
\eta_{j_nki_3\dots i_n}\,\partial_k\,\eta_{j_1j_2\dots j_{n-1} i_2}+
\eta_{j_ni_2ki_4\dots i_n}\,\partial_k\,\eta_{j_1j_2\dots j_{n-1} i_3}
\nonumber\\&\cdots+\eta_{j_ni_2i_3\dots i_{n-1}k}
\,\partial_k\,\eta_{j_1j_2\dots j_{n-1} i_n}-
\eta_{j_1j_2\dots j_{n-1} k}\,\partial_k\,\eta_{j_ni_2i_3\dots i_n}=0.
\label{E:diff}
\end{align}
[In Eq. (5) of \cite{LT} there is a misprint in this formula
(corrected in \cite{CT}): the last term of (\ref{E:diff}) above is
missing.] Note that (\ref{E:brad},\ref{E:ceqd}) is automatically
satisfied for $N\le n+1$ and (\ref{E:diff}) for $N=n$.

Recently it has been shown \cite{Gau} that the algebraic equations
(\ref{E:ceqn},\ref{E:ceqd}) imply that the Nambu tensors are
decomposable (as conjectured in \cite{CT}), which in particular means
that they can be written as determinants of the form
\begin{equation}
\eta_{i_1\dots i_n}=\left|\begin{array}{ccc}
v^1_{i_1}&\dots&v^1_{i_n}\\
\vdots& &\vdots\\
v^n_{i_1}&\dots&v^n_{i_n}\end{array}\right|
=\epsilon_{\alpha_1\dots\alpha_n}
v_{i_1}^{\alpha_1} \cdots v_{i_1}^{\alpha_n}
\label{E:detf}
\end{equation}
In this paper (\ref{E:detf}) is our starting point and we go on studying
the consequences of differential condition (\ref{E:diff}).

\section{Commuting vector fields}
If $\eta$ has the form (\ref{E:detf}) it actually satisfies
(\ref{E:ceqn}) by ${\cal N}=0$ and from this it follows that the
differential equation (\ref{E:diff}) is scale invariant, because for
any scalar $\rho$ we have
\[
{\cal D}_{i_2\dots i_nj_1\dots
j_n}(\rho\eta)= (\rho\partial_k\rho)\,{\cal N}_{ki_2\dots i_nj_1\dots
j_n}(\eta) +\rho^2\,{\cal D}_{i_2\dots i_4j_1\dots j_2}(\eta).
\]
This scale invariance and the determinantal form of $\eta$ imply
certain invariances with respect to changes in the $v$'s, we can
use to define a standard form.  

Let us define an $n\times N$ matrix {\bf V} by $\mbox{\bf V}_{\alpha
k}:=v_k^{\alpha}$, the Nambu tensor $\eta_{i_1\dots i_n}$ is then
given by the determinant consisting of columns $i_1,\dots,i_n$ of
$\mbox{\bf V}$. The rank of {\bf V} must be $n$, otherwise all
$\eta$'s vanish.  If necessary, let us change the numbering so that
the sub-matrix of {\bf V} consisting of its first $n$ columns has
nonzero determinant, and let us denote this $n\times n$ sub-matrix by
$V$. We have $\det V=\eta_{12\dots n}$.

The $n\times N$ matrix $\bar{\mbox{\bf V}}=V^{-1}\mbox{\bf V}$ can now
be used to define another Nambu tensor $\bar\eta$ and we have simply
$\eta=\det V\,\bar\eta$, even if the matrix entries $v$ have changed.
Since in the decomposable case the differential equations are scale
invariant we may equally well consider the Nambu tensor $\bar\eta$.
In this case we have
\begin{equation}
\bar\eta_{12\dots n}=1,\quad 
\bar\eta_{\alpha_1\dots\alpha_{n-1}k}=
\epsilon_{\alpha_1\dots\alpha_{n-1}\alpha_n}\bar v_k^{\alpha_n},\quad
\bar v_i^{\alpha}= \delta_i^\alpha,\mbox{ for }i\le n,\quad
0\le\alpha_k\le n,
\label{E:sta}
\end{equation}
(and correspondingly, $\bar v_k^{\alpha_n}\propto
\epsilon_{\alpha_1\dots\alpha_{n-1}\alpha_n} \bar
\eta_{\alpha_1\dots\alpha_{n-1}k}$). This may be considered the {\em
standard form} for the Nambu tensor and $\bar{\mbox{\bf V}}$ the
standard form of the defining matrix. They are quite useful in
studying the differential part of FI. Furthermore, if the tensor
$\eta$ is given explicitly it may not be so easy to find an equally
simple $\mbox{\bf V}$, but the entries $\bar v_k^{\alpha_n}$ of the
standard form can be read off directly.

Using (\ref{E:detf}) we can write the Nambu bracket (\ref{E:brad})
as
\begin{eqnarray}
\{f_1,\dots,f_n\}&:=&\eta_{i_1\dots i_n}
\partial_{i_1}f_1\cdots\partial_{i_n}f_n
\nonumber\\
&=&\epsilon_{\alpha_1\dots \alpha_n}D^{\alpha_1}f_1\cdots
D^{\alpha_n}f_n\nonumber\\
&=&\rho\epsilon_{\alpha_1\dots \alpha_n}\bar D^{\alpha_1}f_1\cdots
\bar D^{\alpha_n}f_n
\label{E:brad2}
\end{eqnarray}
where
\begin{equation}
D^\alpha:=\sum_{k=1}^N v_k^\alpha\partial_k,\quad
\bar D^\alpha:=\partial_\alpha+\sum_{k=n+1}^N \bar
v_k^\alpha\partial_k,\quad
\rho=\eta_{12\dots n}.
\label{E:Ddef}
\end{equation}
Our main result is the following:

\vskip 0.4cm
\noindent{\bf Theorem:} {\em The $n$'th order Nambu tensor in
dimension $N$, given by (\ref{E:brad},\ref{E:detf}), solves the
differential condition (\ref{E:diff}) iff the differential operators
$\bar D$ of the standard form commute.}

\vskip 0.4cm
\noindent{\bf Proof:} It is clear that if the differential operators
$\bar D$ commute, they behave just like ordinary partial derivatives
in computing the consequences of the FI. As noted before, the overall
factor can be omitted in the decomposable case. Therefore in this case
the Nambu tensor behaves like the canonical one and the conditions
coming from FI are satisfied.

Since the Nambu tensor $\eta$ changes only by an overall factor when
the defining matrix $\mbox{\bf V}$ is multiplied by some matrix $C$
from left, the tensor will continue to satisfy the differential
condition (\ref{E:diff}), even though in other cases the differential
operators might not commute. However, from any given form of $D$'s it
is easy to go to the standard form, and what remains to be proven is
that in that form the differential operators must commute. (If $n=N$
the standard form is the canonical form and there is nothing to
prove.)

In the standard form $[\bar D^\alpha,\bar D^\beta]=0$ is equivalent to
\begin{equation}
\partial_\alpha \bar v^\beta_l+\sum_{k=n+1}^N
\bar v_k^{\alpha}\partial_k \bar v_l^{\beta}=
\partial_\beta \bar v^\alpha_l+\sum_{k=n+1}^N
\bar v_k^{\beta}\partial_k \bar v_l^{\alpha}, \,\forall l>n.
\label{E:pcom}
\end{equation}
Let us now take equation (\ref{E:diff}) for the case when
$j_1,\dots,j_n$ is a permutation of $1,\dots,n$, and
$i_2,\dots,i_{n-1}$ is a permutation of an $n-2$ element subset of
$1,\dots,n$, and $i_n=l>n$ (Here we need $N>n$).  Since $\eta_{12\dots
n}=1$ only the last two terms in (\ref{E:diff}) survive and we get the
condition
\[
\eta_{j_ni_2\dots i_{n-1}k}\,\partial_k\eta_{j_1\dots j_{n-1}l}
=\eta_{j_1\dots j_{n-1}k}\,\partial_k\eta_{j_ni_2\dots i_{n-1}l}.
\]
Contracting this with $\epsilon_{j_ni_2\dots i_{n-1}\alpha}
\epsilon_{j_1\dots j_{n-1}\beta}$ and recalling (\ref{E:sta}) yields
(\ref{E:pcom}).$\Box$

In Theorem 2 of \cite{Gau} similar conclusions are reached but the
approach of that paper is quite different.

\subsection*{Example}
As an example let us consider $n=3,\,N=4$ with $\eta_{ijk}=
\epsilon_{ijkl}x_l$ \cite{CT}. It is easy to see that (when $x_4\neq
0$) in the standard form the matrix {\bf V} is
\[
\bar{\mbox{\bf V}}=\left(\begin{array}{cccc}
1&0&0&-x_1/x_4\\
0&1&0&-x_2/x_4\\
0&0&1&-x_3/x_4\end{array}\right).
\]
and clearly the corresponding differential operators $\bar D^\alpha =
\partial_{x_\alpha} -\frac{x_\alpha}{x_4}\partial_{x_4}$ commute.
Multiplying $\bar{\mbox{\bf V}}$ by $x_4$ produces one more
alternative form $\tilde \eta=x_4^2 \eta =x_4^3\bar\eta$, and the
corresponding vector fields are nothing but angular momentum
operators: $\tilde D^\alpha= L_{4\alpha}= x_4\partial_\alpha -
x_\alpha\partial_4$. Now we have $[\tilde D^\alpha,\tilde
D^\beta]=L_{\alpha\beta}$, but in this form the corresponding vector
fields do not have to commute.

We can complete the analysis of this case by changing into new
variables defined by
\[ 
X_\alpha=x_\alpha,\quad X_4=\tfrac12(x_1^2+x_2^2+x_2^3+ x_4^2)
\]
and correspondingly
\[
\partial_{X_\alpha}=\bar D_\alpha,\quad
\partial_{X_4}=\frac1{x_4}\partial_{x_4}.
\]
All the derivative operators commute, and $\partial_{X_n}\, X_m =
\delta_m^n$.  In the new coordinates the bracket reduces to the
canonical Nambu bracket almost everywhere, that is whenever $x_4\neq
0$. In the omitted subspace one can use some other standard form.
Note that derivatives with respect to the variable $X_4$ do not appear
in the new form of the bracket, $X_4$ is now a constant of motion
(whose form could have seen directly from the given $\eta$). Thus the
motion defined by this $\eta$ takes place on the sphere $
x_1^2+x_2^2+x_2^3+x_4^2 =$const., and its dynamics there is given by
two Nambu Hamiltonians. The fact that the motion takes place on the
surface of a hyper-sphere explains the appearance of angular momentum
operators in the alternative form $\tilde \eta$. There are 6 such
operators but only 3 are needed to move on the surface, the choice
above was to use $L_{\alpha 4}$, which works on the chart where
$x_4>0$ or $<0$.

The above example generalizes immediately to any $N=n+1$: if we take
$\eta_{i_1\dots i_n}= \epsilon_{i_1\dots i_n l}(\partial_l\,m)$, the
motion stays on the surface $m(x_1,\dots,x_{n+1})=$const. That this is
also the most general form for $N=n+1$ (as least locally) can be seen as
follows.  For $N=n+1$ any $\eta$ can be written as $\eta_{i_1\dots
i_n}= \epsilon_{i_1\dots i_n l}f_l$, which means that in the standard
form we have operators $\bar D=\partial_{x_\alpha}
-f_\alpha/f_N\partial_{x_N}$ and their commutation condition is
\begin{equation}
\partial_\alpha g_\beta-g_\alpha\,\partial_N\,g_\beta=
\partial_\beta g_\alpha-g_\beta\,\partial_N\,g_\alpha,
\label{E:34}
\end{equation}
where $g_\alpha=f_\alpha/f_N$. Now let us try to find a function $m$
that solves $ \partial_i m=k f_i,\,i=1,\dots N$. From $i=N$ we get
$k=\partial_N m/f_N$ so that we should solve $\partial_\alpha
m=g_\alpha \partial_N m$ for $\alpha=1,\dots N-1$. The integrability
condition for this set of equations is nothing but (\ref{E:34}). This
means that at least locally we can find the required constant of
motion $m$. Whether this can be done globally is another matter, and
brings in the usual subtleties of chaos vs. integrability.

The next generalization in this direction would be to consider $N=n+2$
with $\eta_{i_1\dots i_n}= \epsilon_{i_1\dots i_n k l}(\partial_k\,f)
(\partial_l\,g)$. Clearly $f$ and $g$ are two conserved quantities in
the corresponding Nambu dynamics. In the standard form we get vector
fields $\bar D^\alpha=\partial_\alpha-\bar
v^\alpha_{n+2}\partial_{n+1}+\bar v^\alpha_{n+1}\partial_{n+2}$
with $\bar v^\alpha_k=(\partial_\alpha f\,\partial_{k} g -
\partial_k f\,\partial_{\alpha} g)
/(\partial_{n+1} f\,\partial_{n+2} g - \partial_{n+2}f\,
\partial_{n+1}g)$, whose commutation can be verified directly.

\section{Nambu tensors from integrable systems}
With the above theorem the problem of constructing Nambu brackets has
been reduced to finding commuting linear differential operators
(\ref{E:Ddef}). A rich set of examples is now provided by integrable
systems.

Let us assume that we have a Liouville integrable system in dimension
$n$, that is we have a set of $n$ functionally independent globally
defined functions in involution, i.e. whose Poisson brackets vanish.
These functions and the underlying Poisson structure define commuting
$2n$-dimensional Hamiltonian vector fields (\cite{Ar}, Sec 8).  With
the canonical Poisson structure the Hamiltonian vector field for a
function $f$ is given by
\[
F:=\sum_{i=1}^n\left(\frac{\partial f}{\partial p_i}\partial_{q_i} -
\frac{\partial f}{\partial q_i}\partial_{p_i}\right),
\]
where $q_i$ and $p_i$ are the canonically conjugate coordinates.
Thus, if $f_i$ are in involution for $i=1\dots n$, then we can define
the matrix elements of $\mbox{\bf V}$ as 
\[
v_i^j=\frac{\partial f_j}{\partial p_i},\quad
v_{i+n}^j=-\frac{\partial f_j}{\partial q_i},\quad\forall 1\le i,j\le n
\]
In this way we get $n$-th order Nambu tensors in dimension $N=2n$.

\subsection{Example}
Let us consider the three-dimensional Toda lattice given by the
Hamiltonian
\[
I_2:=\tfrac12(p_1^2+p_2^2+p_3^2)+e^{q_1-q_2}+e^{q_2-q_3}+e^{q_3-q_1}.
\]
This is integrable, with the other two commuting conserved quantities
given by
\begin{eqnarray}
I_1&:=&p_1+p_2+p_3,\\
I_3&:=&p_1p_2p_3-e^{q_2-q_3}p_1-e^{q_3-q_1}p_2-e^{q_1-q_2}p_3.
\end{eqnarray}
The three commuting vector fields are now
\begin{align}
D_T^1&=\partial_{q_1}+\partial_{q_2}+\partial_{q_3},\\
D_T^2&=p_1\partial_{q_1}+ p_2 \partial_{q_2}+ p_3
\partial_{q_3}+\nonumber\\
&\quad(e^{q_3-q_1}-e^{q_1-q_2})\partial_{p_1}+
(e^{q_1-q_2}-e^{q_2-q_3})\partial_{p_2} +
(e^{q_2-q_3}-e^{q_3-q_1})\partial_{p_3},\\ 
D_T^3&= (p_2p_3-e^{q_2-q_3}) \partial_{q_1}+ 
(p_1p_3-e^{q_3-q_1})\partial_{q_2}+
(p_1p_2-e^{q_1-q_2}) \partial_{q_3}+ \nonumber\\
&\quad(e^{q_1-q_2}p_3-e^{q_3-q_1}p_2 ) \partial_{p_1}+
(e^{q_2-q_3}p_1-e^{q_1-q_2}p_3) \partial_{p_2}+
(e^{q_3-q_1}p_2-e^{q_2-q_3}p_1)\partial_{p_3},
\end{align}
from which the corresponding matrix $\mbox{\bf V}$ can be read.  Since
$D_T^\alpha I_\beta=0$ for all $\alpha,\beta$, the dynamics given by
\[
\dot g =
\{h_1,h_2,g{\}}_T:=\epsilon_{\alpha_1\alpha_2\alpha_3}
(D_T^{\alpha_1}h_1)\,(D_T^{\alpha_2}h_2)\,(D_T^{\alpha_3}g),
\]
has the property that $\dot I_\alpha=0$, no matter what the Nambu
Hamiltonians $h_i$ are. 

Now recall that if an $n$-dimensional Hamiltonian system is Liouville
integrable, then the motion actually takes place on an $n$-dimensional
sub-manifold of the original $2n$-dimensional phase space defined by
$I_i=c_i$, where the constants $c_i$ are determined from the initial
values. The motion on this sub-manifold is still defined by the
original Hamiltonian.

If the dynamics is defined by a Nambu bracket arising from an integrable
system as discussed above, the motion is again restricted to the
manifold defined by $I_i=c_i$, but the motion on this manifold is
now defined by the two additional Nambu Hamiltonians, which we could
choose as we wish.

The other method of using $n$-dimensional integrable systems to define
Nambu dynamics is to use the canonical Nambu tensor of order $2n$ and
the constants of motion as Nambu Hamiltonians, for examples see
\cite{Cha}.

\subsection{Example}
If $N=4$ integrable systems have two commuting quantities, the
Hamiltonian $H$ and a constant of motion $I_2$, but for a third order
Nambu tensor we would need three commuting vector fields. This is
possible in some super-integrable cases, i.e., if we have one more
constant of motion $I_3$. The third constant of motion cannot have a
vanishing Poisson bracket with $I_2$, but the corresponding Hamiltonian
vector fields could still commute. An example is provided by the
following:
\begin{equation}
H=I_1:=F\left((p_1-p_2)^2+(q_1-q_2)^2\right),\quad
I_2:=q_1+q_2,\quad
I_3:=p_1+p_2.
\end{equation}
Now $\{I_2,I_3\}=2$, but the corresponding vector fields
\begin{eqnarray*}
D^1&=&F'\left((p_1-p_2)^2+(q_1-q_2)^2\right)
\left((p_1-p_2)(\partial_{q_1}-\partial_{q_2})-
	(q_1-q_2)(\partial_{p_1}-\partial_{p_2})\right),\\
D^2&=&-(\partial_{p_1}+\partial_{p_2}),\\
D^3&=&\partial_{q_1}+\partial_{q_2},
\end{eqnarray*}
do commute. In this case the standard form of the matrix giving the
$\eta$'s is quite simple, the first three columns form a unit matrix
and the fourth column is given by $\bar v^1_4=-\bar v^2_4=(q_1-q_2)/
(p_1-p_2),\, \bar v^3_4=1$.  In the new variables
$X_1=q_1,X_2=q_2,X_3=p_1,X_4=\frac12(q_1-q_2)^2+ \frac12 (p_1-p_2)^2$,
$\partial_{X_\alpha}=\bar D^\alpha,$ for $\alpha=1,2,3$,
$\partial_{X_4}=1/(p_1-p_2)\partial_{p_2}$ the Nambu bracket reduces
to the canonical one and the motion stays on the manifold $X_4=$const.
(In \cite{CT} the same functions were used as Hamiltonians in a fourth
order canonical Nambu bracket in dimension four.)

Another super-integrable example but with non-algebraic constants of
motion is given by \cite{JH}:
\[
H=I_1:=\tfrac12 p_1^2+\tfrac12 p_2^2 -p_2q_1/q_2,\quad
I_2:=(q_1 p_2-q_2 p_1+q_2)/p_2,\quad
I_3:=p_1+\log(p_2/q_2).
\]
Again  $\{I_2,I_3\}=1$, but the Hamiltonian vector fields
\begin{eqnarray*}
D^1&=&p_1\partial_{q_1}+(p_2-q_1/q_2)\partial_{q_2}+(p_2/q_2)\partial_{p_1}-
(p_2q_1/q_2^2)\partial_{p_2},\\
D^2&=&-(q_2/p_2)\partial_{q_1}+(q_2(p_1-1)/p_2^2)\partial_{q_2}
-\partial_{p_1}+((p_1-1)/p_2)\partial_{p_2},\\
D^3&=&\partial_{q_1}+(1/p_2)\partial_{q_2}+(1/q_2)\partial_{p_2}.
\end{eqnarray*}
commute. In the standard form the last column of $\bar{\mbox{\bf V}}$
is given by $\bar v^1_4=p_2/(p_2q_2-q_1),\, \bar
v^2_4=-p_2q_1/(q_2(p_2q_2-q_1)),\, \bar v^3_4=-p_1q_2/(p_2q_2-q_1).$
The Hamiltonian defines a (non-compact) manifold on which the motion
takes place, and the new variables on which the Nambu tensor is
canonical are $X_1=q_1,X_2=q_2,X_3=p_1,X_4=\tfrac12 p_1^2+\tfrac12
p_2^2 -p_2q_1/q_2$,with $\partial_{X_\alpha}=\bar D^\alpha,$ for
$\alpha=1,2,3$, and $\partial_{X_4}=q_2/(p_2q_2-q_1)\partial_{p_2}$.

\section{Discussion}
In this letter we have studied the differential part (\ref{E:diff}) of
the Fundamental Identity on the assumption that the algebraic part
implies decomposability (\ref{E:detf}). A standard form
(\ref{E:sta}-\ref{E:Ddef}) has been defined for the Nambu tensor in this
case and the differential condition was related to the commutativity
of the corresponding vector fields $\bar D$.

The simplest Nambu tensor of order $n$ is obtained in dimension $N=n$
and is given by the totally antisymmetric constant tensor. The present
results indicate, that the dynamics defined by a Nambu bracket of the
same order but in higher dimensions is still essentially
$n$-dimensional.

If we define the Nambu bracket using the Hamiltonian vector field of a
Liouville integrable system, then the bracket itself guarantees that the
motion stays on the $n$-dimensional manifold defined by the constants
of motion of the underlying integrable system. The motion on this
manifold is determined by the Nambu Hamiltonians $h_i$, and this
motion does not have to be integrable.

One important open problem has been the quantization of the dynamics
defined by a Nambu bracket. The connection to integrable systems via
commuting vector fields presented in this letter will hopefully bring
new light to this question as well.

\section*{Acknowledgments}
I would like to thank L. Takhtajan for discussions and for comments on
the manuscript.  This work was supported in part by the Academy of
Finland, project 31445.

\end{document}